\documentclass[fleqn,usenatbib]{mnras}

\usepackage{newtxtext,newtxmath}

\usepackage[T1]{fontenc}
\usepackage{ae,aecompl}
\usepackage{tablefootnote}   
\usepackage{booktabs,caption,fixltx2e}
\usepackage[flushleft]{threeparttable}

\usepackage{graphicx}	
\usepackage{amsmath}	
\usepackage{amssymb}	

\title[Gamma-ray quasi-periodicities of blazars]{
Gamma-ray quasi-periodicities of blazars. A cautious approach
}

\author[S. Covino, A. Sandrinelli \& A. Treves ]{
S. Covino$^{1}
$\thanks{E-mail:--------- }, 
A. Sandrinelli$^{1,2}$ 
and A. Treves$^{1,2}$\\
\\
$^{1}$INAF -  Istituto Nazionale di Astrofisica, Osservatorio Astronomico di Brera, Via Emilio Bianchi 46, I-23807 Merate, Italy\\
$^{2}$Universit\`a degli Studi dell'Insubria, Via Valleggio 11, I-22100 Como, Italy\\
}

\date{Accepted XXX. Received YYY; in original form ZZZ}

\pubyear{2018}

\begin{document}
\label{firstpage}
\pagerange{\pageref{firstpage}--\pageref{lastpage}}
\maketitle

\begin{abstract}
The availability of about a decade of uninterrupted sky monitoring by the \textit{Fermi} satellite has made possible to
study long-term quasi-periodicities for high-energy sources. It is therefore not a surprise that for several blazars
in the recent literature claims for such periodicities, with various level of confidence, have been reported. The confirmation of these findings could be of tremendous importance for the physical description of this category of sources
and have consequences for the gravitational wave background interpretation.
In this work we carry out a temporal analysis of the \textit{Fermi} light curves for several of the sources mentioned in recent
literature by means of a homogeneous procedure and find that, globally, no strong cases for blazar year-long quasi-periodicities can be confirmed. The computed power spectral densities are all essentially consistent with being generated by red-noise only. We further discuss the meaning and the limitations of the present analysis.
\end{abstract}




\section{Introduction}

Blazars are the dominant sources of the extragalactic $\gamma$-ray sky. 
The \textit{Fermi} mission, with its continuous monitoring of the entire sky,
produced light curves of the brightest objects for a duration now approaching a decade and,
by adopting a week/month binning, the curves are also practically evenly sampled. 
As in other spectral bands the light curves indicate large variability basically on any time scale. 
A topic of particular interest is about the possibility to identify a periodic behavior superposed 
to the usually dominant stochastic variability. 
Long periods, of the order of months or years, are of particular relevance, since the merger
of two supermassive black holes (M$\sim$10$^8$ M$_{\odot}$), possibly the final act of the interaction 
of two galaxies, could result in year-long orbital periods \citep[e.g.][]{Begelmanetal1980}.
As a matter of fact, with year-long periods, the number of full cycles covered 
by the \textit{Fermi} monitoring is unavoidably small, and this clearly affects the estimates of 
the significance of any claimed periodicity and increases the probability of spurious detections. 
 
In the recent literature several claims for year-long periodicities based on analyses of \textit{Fermi} data of 
have been proposed \citep[i.e.][]{Sandrinellietal2014,Ackermannetal2015,Sandrinellietal2016a,Sandrinellietal2016b,ProkhorovMoraghan2017,Sandrinellietal2017,Covinoetal2017,Zhangetal2017a,Zhangetal2017b,Zhangetal2017c,Tavanietal2018,Sandrinellietal2018}. These analyses rely on different procedures and assumptions, making often a direct comparison arduous.
In this paper we therefore reanalyze the gamma-ray light curves for the objects reported in literature following
a homogeneous procedure well suited for relatively high signal to noise (S/N) evenly sampled light curves, and critically 
evaluate the solidity of these claims. In Sect.\,\ref{sec:search} we describe the analysis procedure we have applied, in Sect.\,\ref{sec:sample} we present the \textit{Fermi} blazar sample we have selected, 
in Sect.\,\ref{sec:results} we describe and discuss our main results and, finally, in Sect.\,\ref{sec:discconcl}, we summarize 
our conclusions.

\section{Data and Methods}
\label{sec:search}
 
We considered aperture photometry light curves  with 30 day time resolution in the 100\,MeV to 200\,GeV energy range available from the Fermi website\footnote{https://fermi.gsfc.nasa.gov/ssc/data/access/lat/4yr\_catalog/ap\_lcs.php}. They cover about a decade of uninterrupted observations for all the sources considered in this work. The regular sampling considerable simplifies the timing analysis.

Blazar light curves are characterized by intense variability at any time scale and show evidence of correlated noise \citep[e.g.][]{Press1978,Abdoetal2010,Lindforsetal2016,Goyaletal2017}. Singling out possible quasi-periodicities in these data is a difficult task and requires to model the red noise in order to derive an assessment of the significance of any possible periods with respect to the noise model. Red noise can also, if not properly modeled, alter the output of the analysis due to leakage of power from lower to higher frequencies possibly and mimicking unreal quasi-periodicities often of transient nature \citep{Kellyetal2014}. We mainly followed the procedure described in \citet{Vaughan2010,Vaughan2013} and \citet{Guidorzietal2016}. Power Density Spectra (PDS) are derived by discrete Fourier transform and are normalized following the considerations reported in \citet{Leahyetal1983} and \citet{Guidorzi2011}. The noise was modeled as a power-law (PL) plus a constant \citep[e.g.][]{Konig&Timmer1997,Kellyetal2009,Edelsonetal2013}:
\begin{equation}
S_{\rm PL} = Nf^{-\alpha}+G,
\end{equation}
where $N$ is a normalization factor, $f$ the frequency, $\alpha$ the PL index and $G$ is the uncorrelated statistical noise that has a value of 2 for pure Poissonian noise with the adopted normalization \citep[see also][]{Guidorzietal2016}. We computed the best-fitting parameters for our PDS in a Bayesian framework. We initially maximized the \textit{Whittle} likelihood function \citep{Barret&Vaughn2012} by a non-linear optimization algorithm and integrate the posterior probability density of the parameters of our models by a Markov Chain Monte Carlo (MCMC) affine-invariant Hamiltonian algorithm \citep{Foreman-Mackeyetal2013}. We started the chains from small Gaussian balls centered on the best fit values. The first third of each chain (the ``burn-in phase") was discarded and we checked that a stationary distribution was reached \citep{Sharma2017}.
Fit quality was evaluated by Kolmogorov-Smirnov (KS) tests of the residuals against the expected $\chi^2$ distribution with two degrees of freedom and from posterior predictive assessment \citep{Gelmanetal1996} based on the ``summed square error" test statistics \citep[TSSE,][]{Vaughan2010}.
Posterior predictive assessment is a Bayesian technique to evaluate any test statistics over the range of parameter values, weighted by the posterior distribution of parameters \citep[see also, e.g.][]{Protassovetal2002,Lucy2018}.
We assumed a Jeffrey prior for the normalization, strict positivity for the PL index, and a large Gaussian distribution centered on 2 for the uncorrelated noise. In Bayesian analysis, the posterior distribution is a complete summary of our inference about the parameters given the data, model and any prior information. For each set of simulated parameters, we generated spectral models and use this to generate a periodogram from the posterior predictive distribution.
Then, sampling the parameters from the posterior distribution, we drew the $T_{\rm R} = \max_j R_j$ statistics to evaluate the global significance of any peak in the PDS \citep[see][for more details]{Vaughan2010,Guidorzietal2016}, where $R = 2P/S$, $P$ is the simulated or observed PDS, and $S$ the best-fit PDS model. This statistic selects the maximum deviation from the continuum spectrum for each simulated PDS. The observed value of $T_{\rm R}$ is then compared with the simulated distribution and the significance is evaluated directly. Given that the same procedure is applied to the simulated as well as to the real data a correction for the multiple trials carried out is already included in the analysis.

\section{The \textit{Fermi} blazar Sample}
\label{sec:sample}
Our sample consists of ten objects: PKS\,0301$-$243, PKS\,0426$-$380, PKS\,0537$-$441, S5\,0716$+$714, PKS\,0805$-$077, 4C\,$+$01.28, PG\,1553$+$113, PKS\,2052$-$474, PKS\,2155$-$304 and BL\,Lac. These sources have been selected since a possible periodicity in their \textit{Fermi} light-curves was proposed in the literature. No attempt to secure a complete sample based on any criterion was tried. With some more detail, for PKS\,0301$-$243 a periodicity of about 2.1\,years was suggested by \citet{Zhangetal2017c}. An even longer periodicity of about 3.4\,years was proposed for PKS\,0426$-$380 by \citet{Zhangetal2017b}. S5\,0716$+$714 was studied in \citet{Sandrinellietal2017} and no interesting periodicities were proposed. However, S5\,0716$+$714 ($\sim 346$\,day), PKS\,0805$-$077 ($\sim 658$\,days), 4C\,$+$01.28 ($\sim 445$\,days), PG\,1553$+$113 ($\sim 798$\,days, PKS\,2052$-$474 ($\sim 637$\,days), PKS\,2155$-$304 ($\sim 644$\,days) and BL\,Lac ($\sim 698$\,days) are part of a set of blazars suggested to show periodical behavior in their high-energy data by \citet{ProkhorovMoraghan2017}. Some of these sources are among the most extensively studied blazars at any wavelength. In particular, for PKS\,0537$-$441, \citet{Sandrinellietal2016b} singled out a $\sim 280$\,day periodicity analyzing the high-energy data during a high-state period. A periodicity at about 2.2\,years for PG\,1553$+$113 was initially proposed by \citet{Ackermannetal2015} and then discussed in several more studies, i.e. \citet{Stamerraetal2016,Cutinietal2016,Sandrinellietal2018,Tavanietal2018}. For PKS\,2155$-$304 a periodicity of about 642\,days was suggested by \citet{Sandrinellietal2014,Sandrinellietal2016a,Zhangetal2017a,Sandrinellietal2018}. Finally, BL\,Lac was discussed in \citet{Sandrinellietal2018} and a periodicity of about 680\,days was proposed.

In a few cases, the reported claims only mention hint of periodicities, the significancies are modest, although multiwavelength studies could improve the solidity of these detections (e.g. for PKS\,0537$-$441, PG\,1553$+$113, PKS\,2155$-$304 and BL\,Lac).

\begin{figure*}
\begin{tabular}{cc}
\includegraphics[width=6cm]{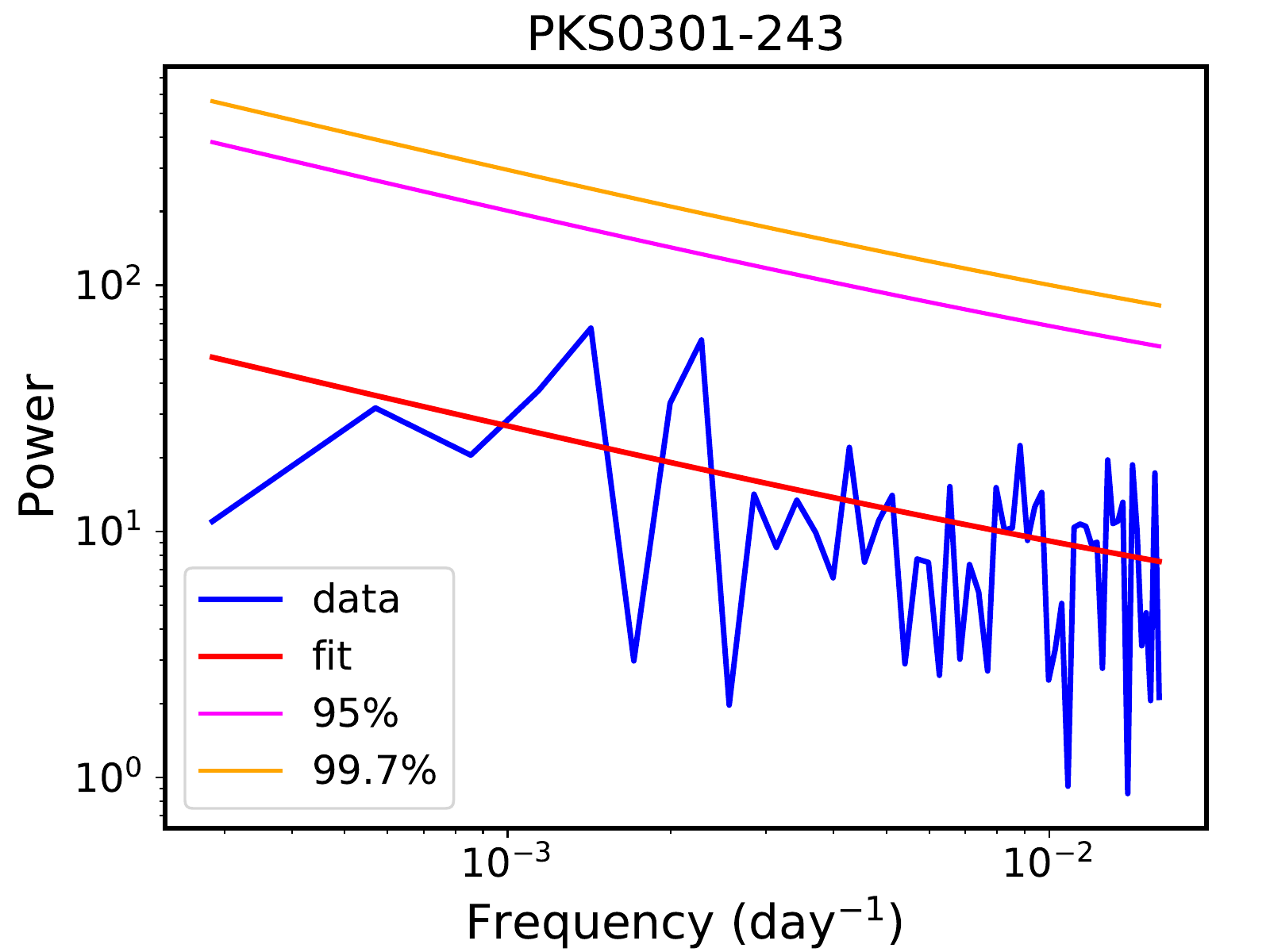} & \includegraphics[width=6cm]{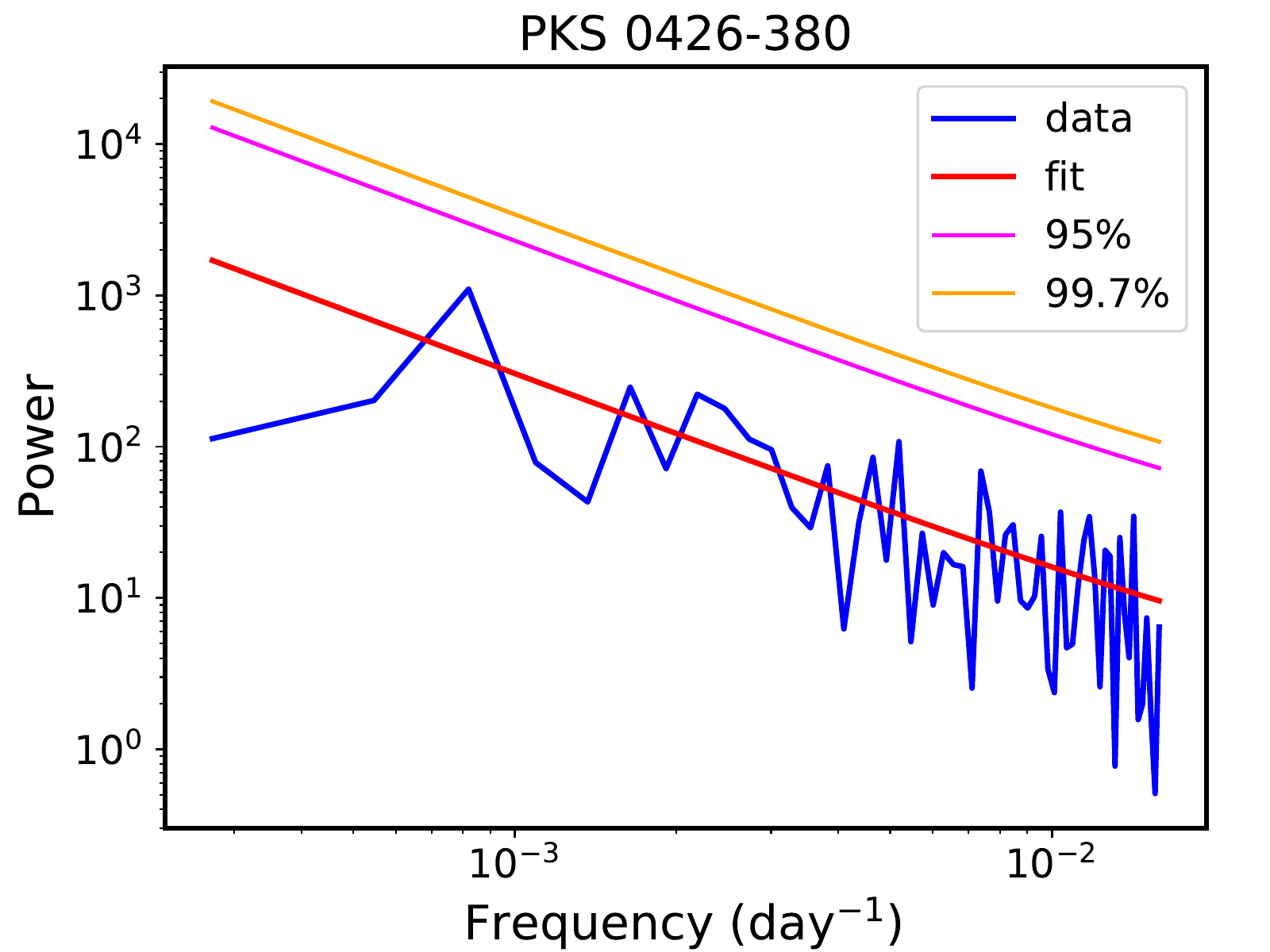} \\
\includegraphics[width=6cm]{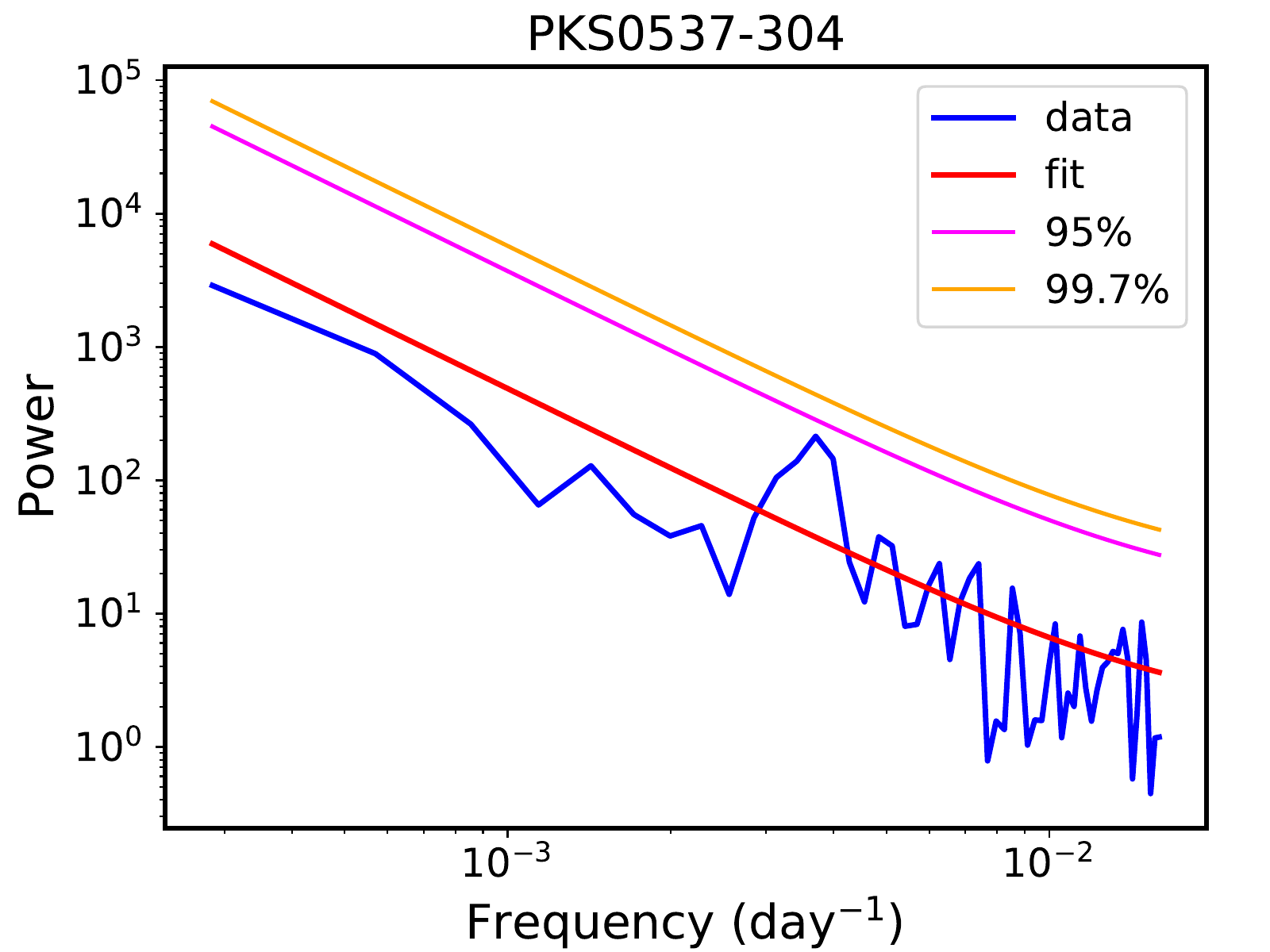} &  \includegraphics[width=6cm]{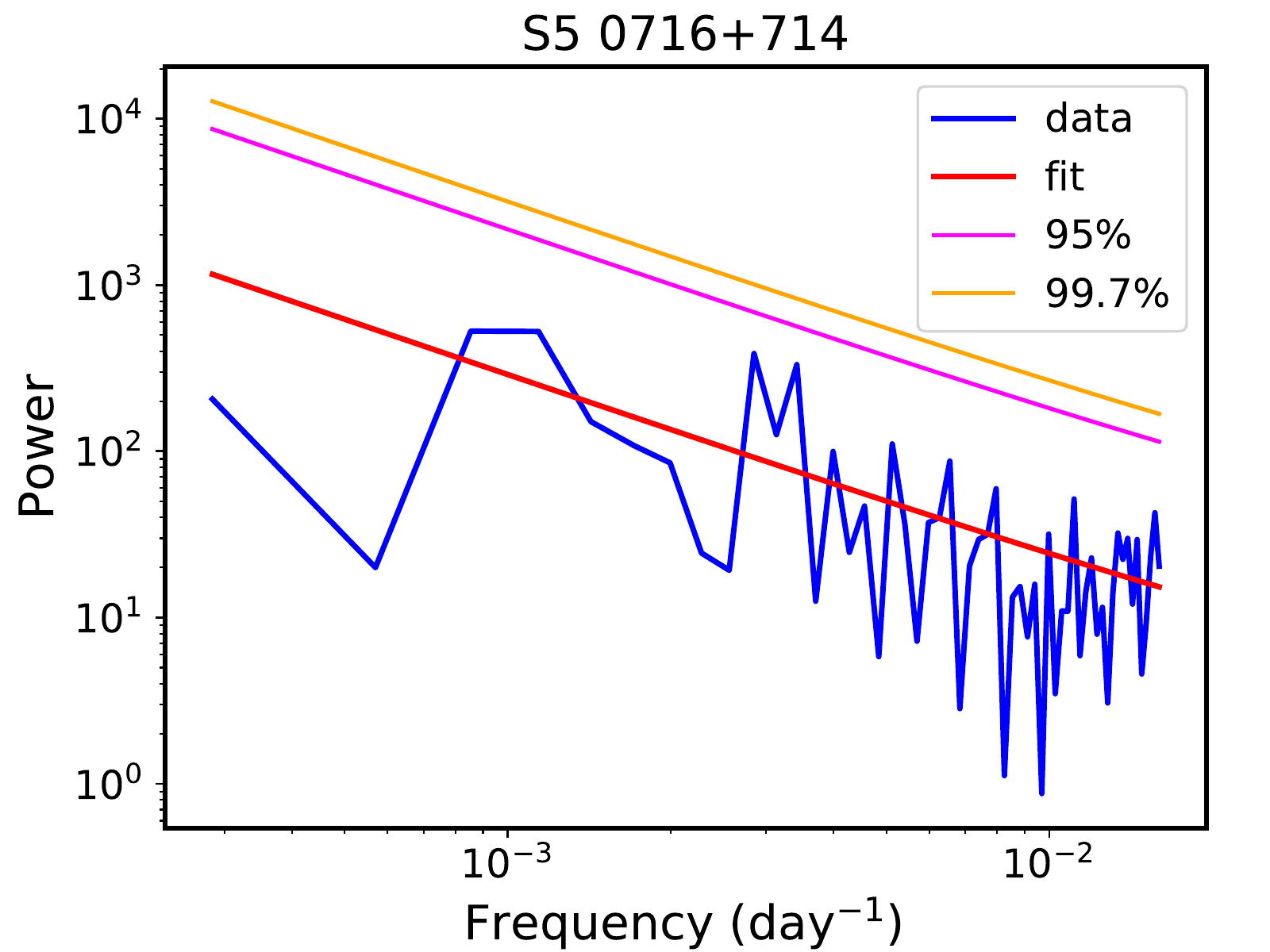} \\
\includegraphics[width=6cm]{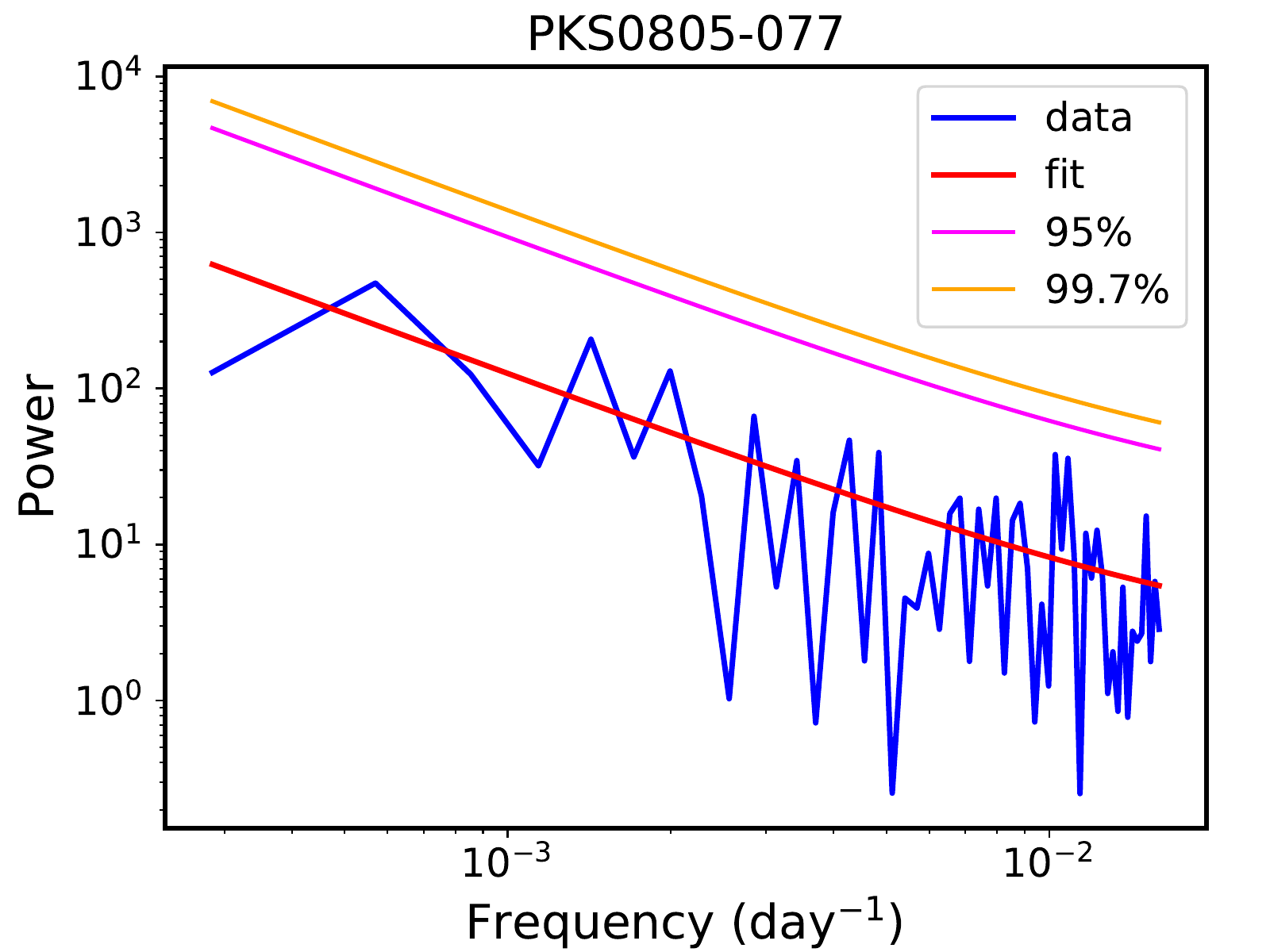} & \includegraphics[width=6cm]{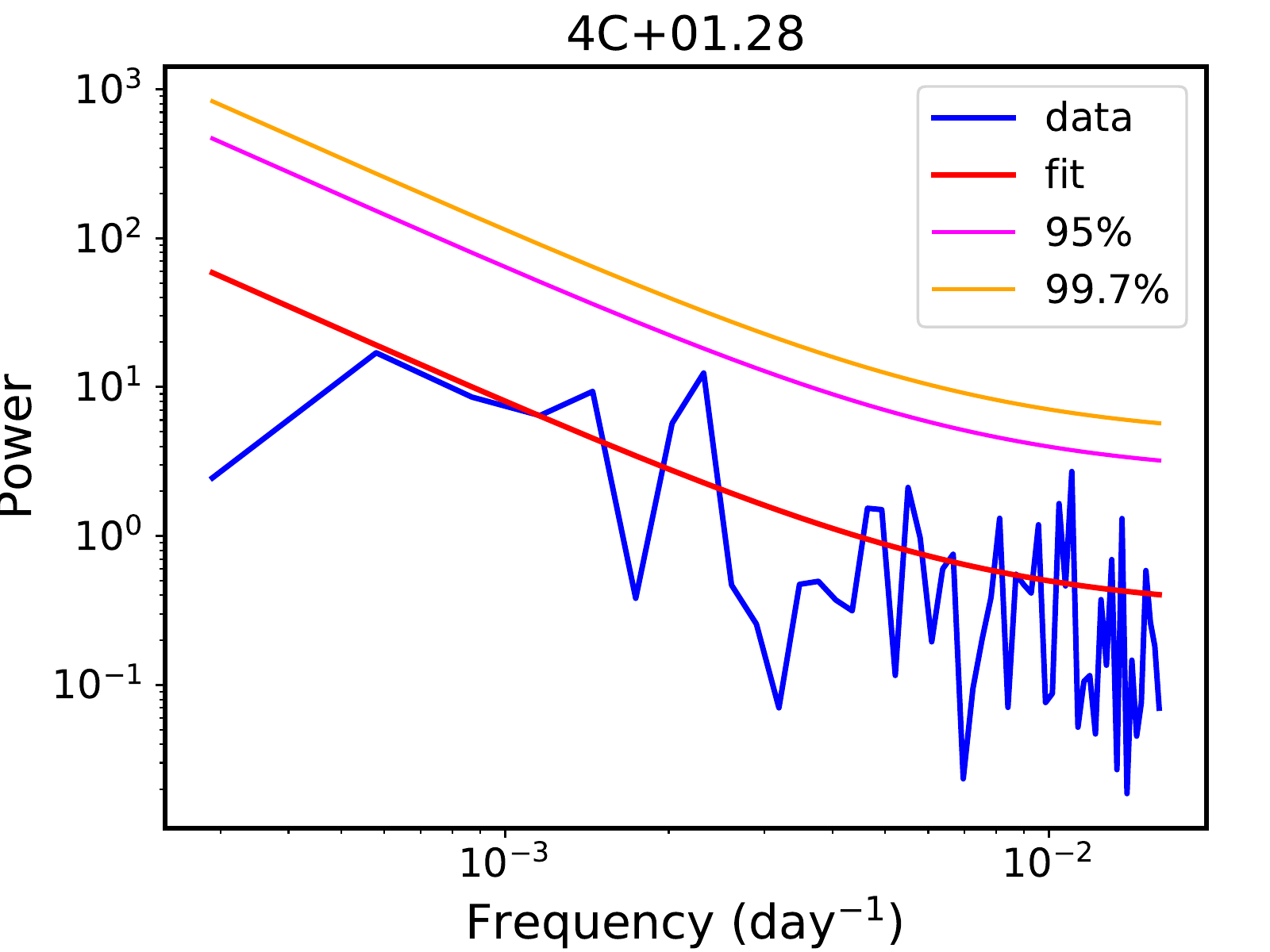} \\
\includegraphics[width=6cm]{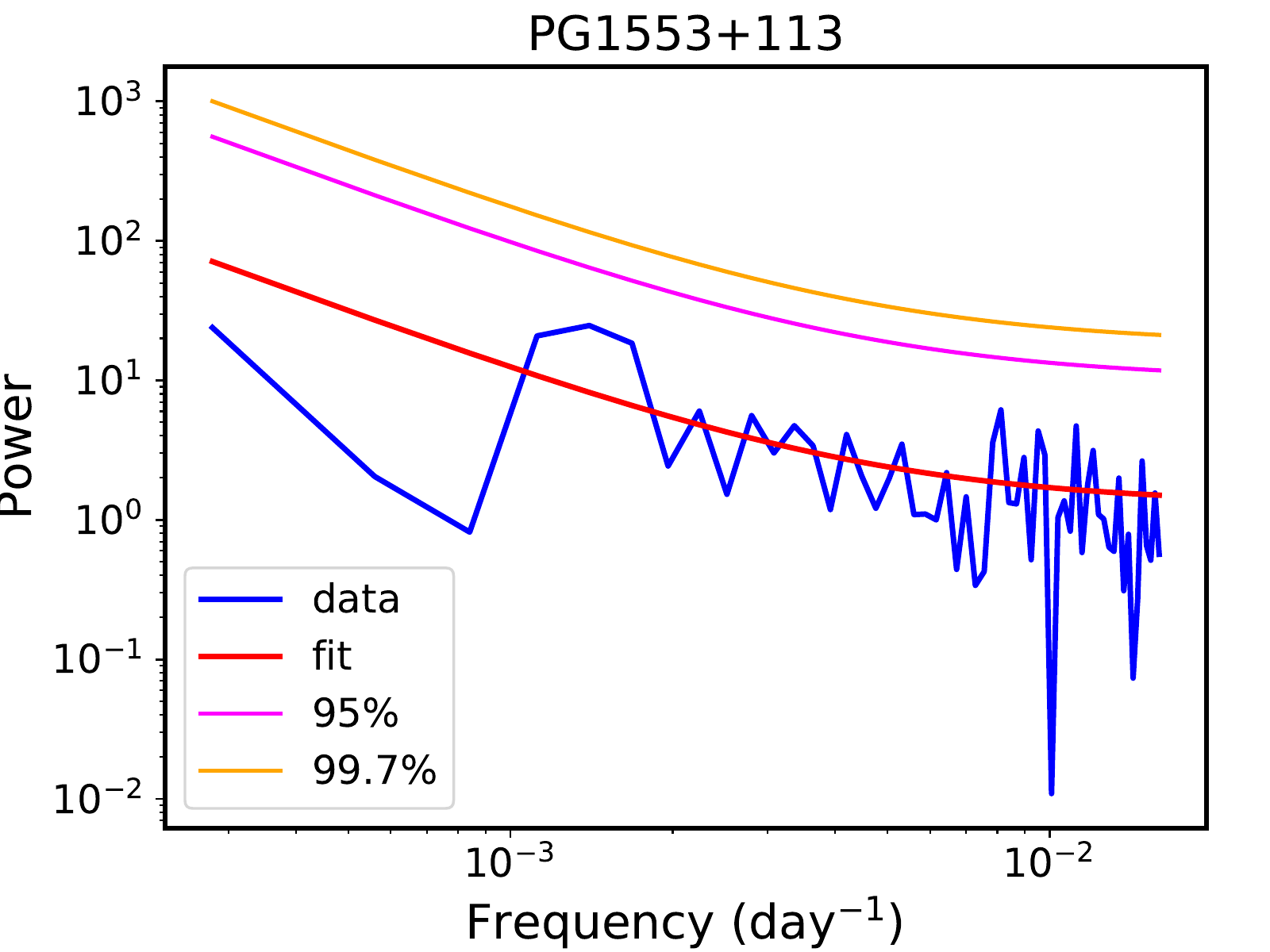} & \includegraphics[width=6cm]{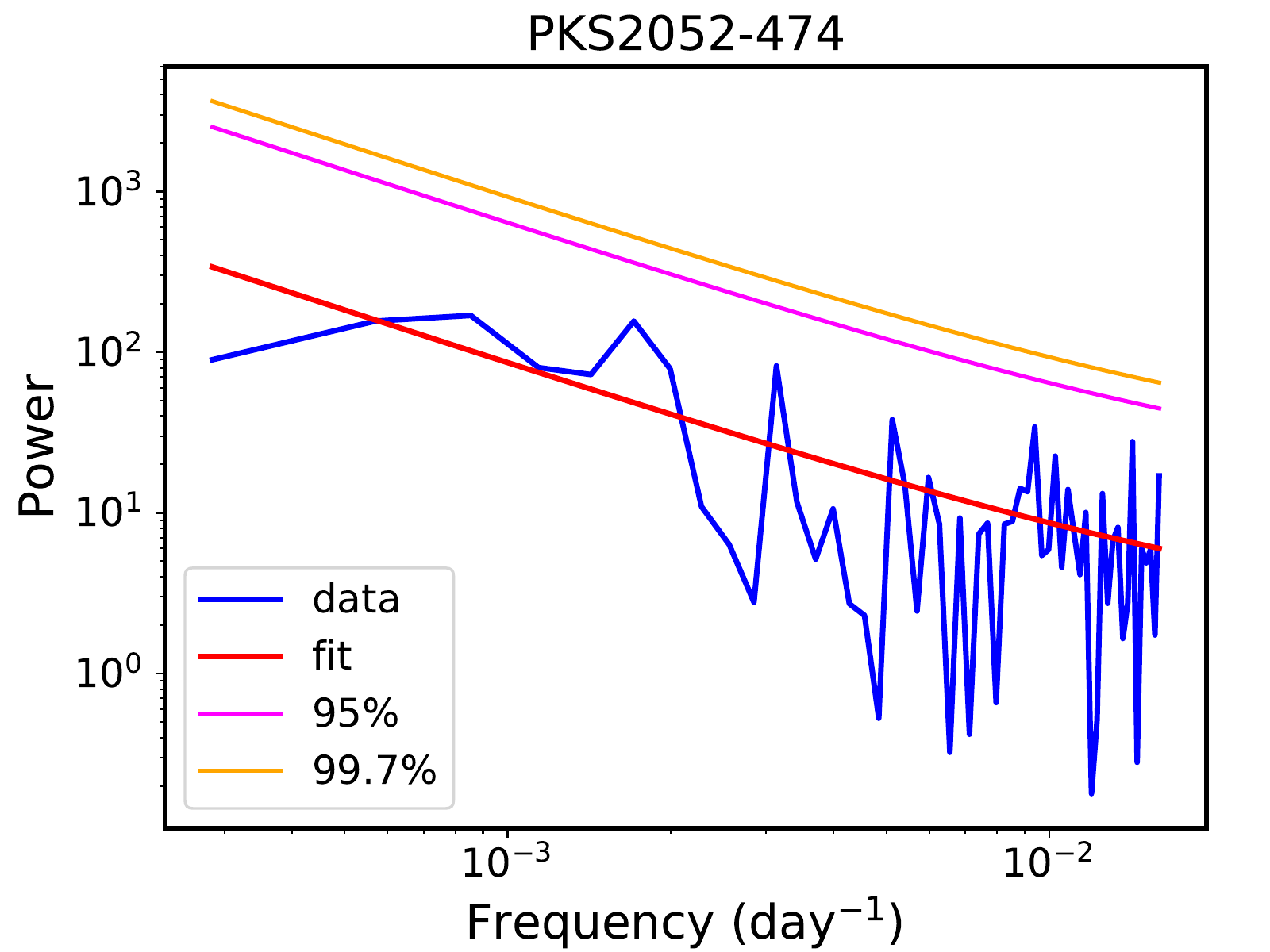} \\
\includegraphics[width=6cm]{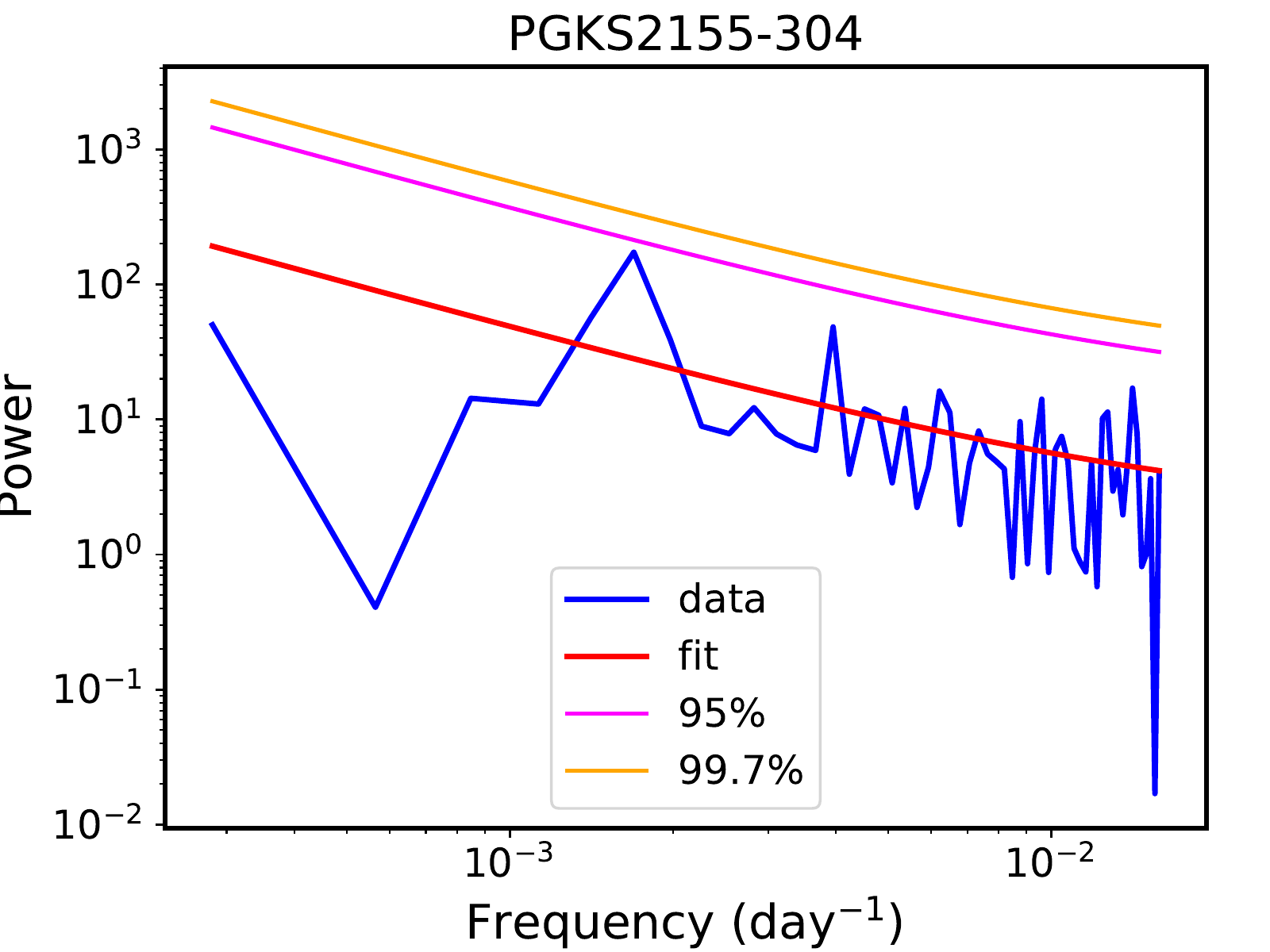} &  \includegraphics[width=6cm]{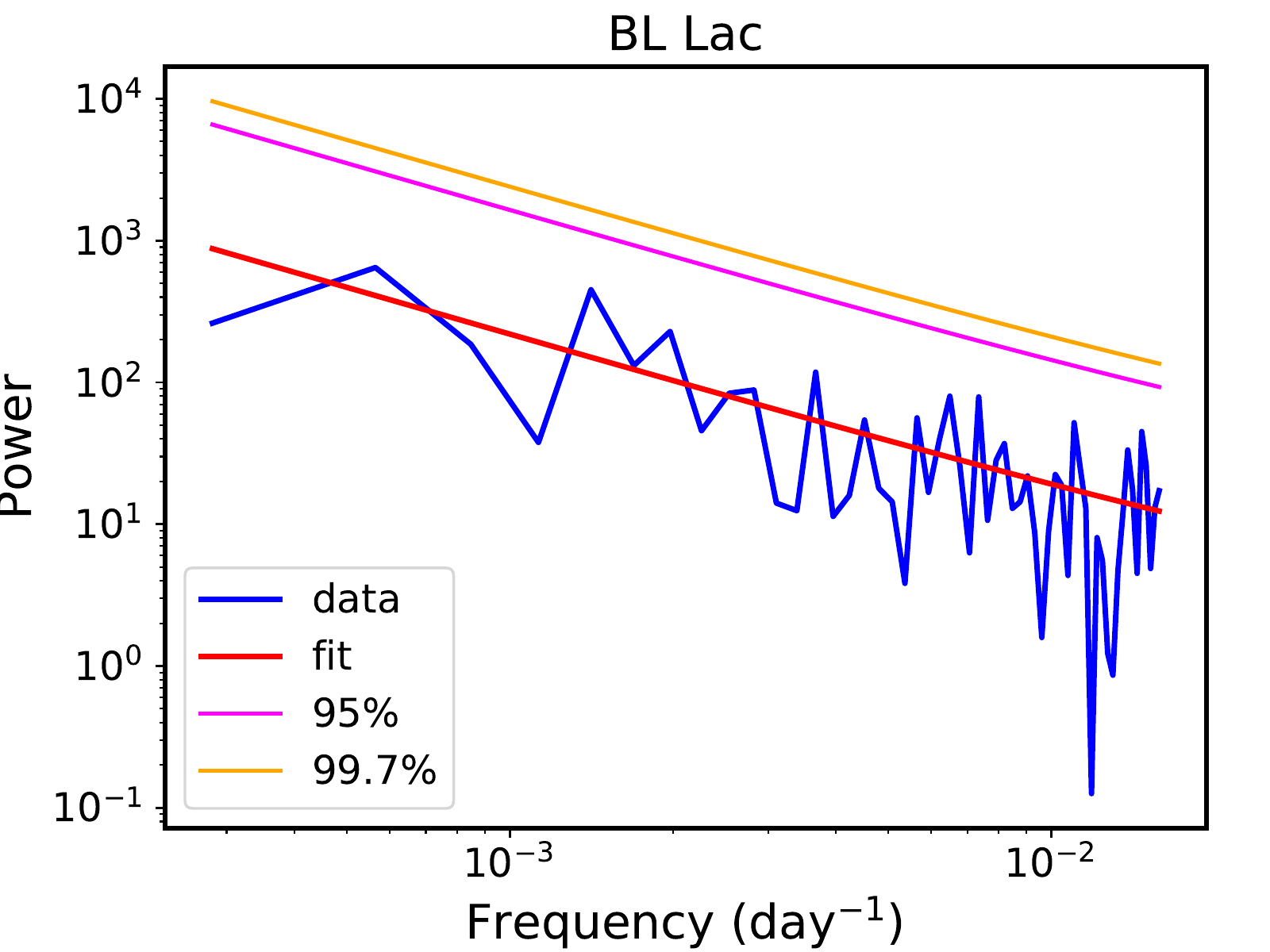} 
\end{tabular}
\caption{PDS (blue) for the blazars considered in this study. The best fit noise model (red). The 95\% (magenta) and
99.7\% (orange) limits are computed evaluating the global significance of any peak in the PDS (see Sect.\,{sec:search}).}
\label{fig:pds}
\end{figure*}

\section{Results and Discussion}
\label{sec:results}
Our results can simply be summarized stating that all the analyzed PDS (see Fig.\ref{fig:pds}) are compatible with being due to noise only, i.e. no periodicities can be singled out at a significance level better than 95\%. In Table\,\ref{tab:slopes} we also report the PDS best fit parameters. In a fair fraction of cases the PL index is consistent with $\sim 1$, although there are a few exceptions showing a steeper relation as for PKS\,0537$-$441. Features (maxima in the computed PDS) of some interest can anyway be identified in the periodograms as, e.g. for PKS\,2155$-$304 or PKS\,0537$-$441, and typically they correspond to the features already proposed in the literature for these sources.

\begin{table}
\centering
\begin{tabular}{lcccc}
\hline
Source & PL index & Poisson noise & $KS$ & TSSE \\
 & & & p-value & p-value \\
\hline
PKS\,0301$-$243 & $0.54^{+0.13}_{-0.13}$ & $2.03^{+0.40}_{-0.42}$ & 0.24 & 0.05 \\
PKS\,0426$-$380 & $1.33^{+0.13}_{-0.12}$ & $1.96^{+0.40}_{-0.41}$ & 0.80 & 0.26 \\
PKS\,0537$-$441 & $2.00^{+0.17}_{-0.15}$ & $1.68^{+0.39}_{-0.39}$ & 0.55 & 0.66 \\
S5\,0716$+$714 & $1.11^{+0.13}_{-0.12}$ & $2.01^{+0.40}_{-0.40}$ & 0.95 & 0.54 \\
PKS\,0805$-$077 & $1.29^{+0.13}_{-0.12}$ & $2.02^{+0.39}_{-0.39}$ & 0.82 & 0.66 \\
4C\,$+$01.28 & $1.63^{+0.32}_{-0.27}$ & $0.32^{+0.13}_{-0.11}$ & 0.03 & 0.89 \\
PG\,1553$+$113 & $1.44^{+0.31}_{-0.25}$ & $1.29^{+0.26}_{-0.26}$ & 0.47 & 0.15 \\
PKS\,2052$-$474 & $1.11^{+0.14}_{-0.12}$ & $2.08^{+0.38}_{-0.39}$ & 0.99 & 0.55 \\
PKS\,2155$-$304 & $1.11^{+0.17}_{-0.15}$ & $1.96^{+0.39}_{-0.38}$ & 0.97 & 0.75 \\
BL\,Lac & $1.10^{+0.12}_{-0.11}$ & $2.02^{+0.42}_{-0.40}$ & 0.78 & 0.26 \\
\end{tabular}
\caption{PDS best fit parameters for the noise model adopted in the present analyses. 1$\sigma$ credible intervals are also reported. The reported p-values are computed by a KS test with respect to the theoretical $\chi^2$ distribution with two degrees of freedom and by posterior predictive assessment \citep{Gelmanetal1996} based on the TSSE test statistics.}
\label{tab:slopes}
\end{table}


The lack of statistically solid quasi-periodicities in the sample can be partly surprising due to the rich literature about this subject. However, in several cases, the authors only mention hints of periodical behaviors, and often based on the coincidence of modest-significance features in PDS computed for light-curves in different bands \citep[e.g.][and references therein]{Sandrinellietal2018}. The present study typically agrees with these past results for what concerns the analysis of the high-energy light curves. In other cases, differences in the adopted analysis techniques, or even the omission of the multi-trial correction, can probably partly justify the different results. It is not expected, in general, that highly significant features can be lost adopting different, well posed, analysis recipes. 

Our analysis is rather standard but it is unavoidable based on several assumption that it is worth reminding here \citep[see also][ for a more comprehensive discussion]{Vaughan2010}. First of all the \textit{Whittle} likelihood is known to be only an approximation of the true sampling distribution of the periodograms, in particular for light-curves of moderate length. A probably much more important concern is due to the nature of the computed periodograms that, being based on decompositions of the time series onto an ortho-normal basis formed by sinusoidal functions, is clearly more sensitive to oscillations following a sinusoidal pattern \citep[see also][]{vanderKlis1989}. This might be particularly important for sources the periodicity of which are inferred by the recurrence of flare-like events more than a real long-lasting modulation. It is also worth to remember here that blazar light curves  are severely dominated by stochastic variability. Non-parametric analyses \citep[e.g.][]{Huijseetal2018,Tavanietal2018} can likely alleviate this difficulty. Our analysis is also performed at the Fourier frequency grid that might not be necessarily the most suitable for identifying possible periodicities. In any case, as already mentioned, looking for possible periodicities of at least several months (bin time 30\,days), implies that only a few cycles could have been effectively covered by the \textit{Fermi} monitoring, making the evaluation of the statistical weight of the possibly identified features intrinsically more difficult. None of the mentioned limitations are actually expected to be able to hidden very significant long periodicities, while they can definitely give negative results in case of modest significance features, although a discussion of the many pros and cons of different recipes for temporal analysis is well beyond the scope of this work.

One more point worth mentioning is not related to the temporal analysis, but to the \textit{Fermi} light-curves adopted in this paper. They are those provided automatically by the \textit{Fermi} team and suffer from a few limitations since they are not background subtracted and are possibly contaminated by nearby bright sources. While these are important details to consider, the sources selected in this work are all rather bright and, for a periodicity analysis, the reported limitations should not constitute a real issue.

\section{Conclusions}
\label{sec:discconcl}
The main result of our study is that there is no solid evidence for year-long periodicities based on the homogeneous procedure applied to the high energy emission data available  at this time for several 
$\gamma$-ray bright blazars. Considering the intrinsic complexity of a temporal analysis, the limited analyzed sample, and the role, often underestimated, of the assumptions behind any recipe, there is still room for single events showing moderately significant oscillations. There is little doubt that this phenomenon is anyway not common for the whole population of blazars. In particular the estimate of $\sim 10$\% of bright gamma-ray blazars exhibiting year-long periodicities proposed by \citet{Sandrinellietal2017,Sandrinellietal2018}, which was based on the results of \citet{ProkhorovMoraghan2017} appears now poorly justified.

Nevertheless, identifying a solid periodicity for a blazar could have an important impact. For instance, adopting the most popular interpretative scenario, a year-long periodicity could be due to supermassive black holes forming a binary system \citep[e.g.][]{Lehto&Valtonen1996,Grahametal2015}. This scenario has indeed been recently invoked and discussed by several authors \citep[e.g.][]{Sandrinellietal2014,Ackermannetal2015,Sandrinellietal2016a,Sobacchietal2017,Cavaliereetal2017,Capronietal2017,Tavanietal2018,Yanetal2018} for some of the blazars considered in this study. A direct consequence of a relatively large population of binary supermassive black holes could be the emission of gravitational waves producing a background important in the frequency range covered by the Pulsar Timing Array \citep{Hobbs&Dai2017}. As discussed in \citet{Sandrinellietal2018} and \citet{Holgadoetal2018}, already now the upper limits constrain the possible binary fraction to be much lower than 10\% of the whole population, thus indirectly confirming our negative results.

The low fraction of systems showing year-long quasi-periodic oscillations also constrain other scenarios where the oscillations are induced 
by instabilities in the relativistic jet or in the accretion disk \citep[e.g.][]{Camenzind&Krockenberger1992,Marscher2014,Raiterietal2017} requiring it is again relatively uncommon, although it is difficult to derive strict upper limits on the presence of quasi-periodic oscillations from our sample due to the biased selection.

It is anyway clear that the availability of high-quality and well sampled light curves in the $\gamma$-rays open exciting perspectives for timing analysis both to confirm low-significance yet not negligible periodicity already proposed and to sample time-scales still not accessible due to the (relatively) limited length of the \textit{Fermi} monitoring. We finally mention that interesting perspectives for singling out binary supermassive black-holes are also allowed by the future availability of accurate astrometric measurements by the GAIA\footnote{http://sci.esa.int/gaia/} mission as discussed in \citet{DOrazio&Loeb2018}. This technique could offer a synergistic view to the problem since it does not require to solve the difficult problem to identify possible periodicities in their electromagnetic emission hidden in the always present stochastic variability.

\section*{Acknowledgements}
We acknowledge the anonymous referee for her/his useful comments and suggestions that greatly improved the readability of the paper.
We acknowledge partial funding from Agenzia Spaziale Italiana-Istituto Nazionale di Astrofisica grant I/004/11/3. We also thank dr. Filippo D'Ammando for useful discussions and suggestions about the {\textit Fermi} data products.




\bibliographystyle{mnras}
\bibliography{qpo} 

\bsp	
\label{lastpage}
\end{document}